\documentclass[10pt]{iopart}
\usepackage{iopams}
\usepackage{amsfonts,amssymb,graphicx}
\usepackage{color}
\usepackage[dvips]{hyperref}

\newcommand{\bfr}{{\bf r}}
\newcommand{\bfp}{{\bf p}}
\newcommand{\bfq}{{\bf q}}
\newcommand{\lambdab}{\widetilde{\lambda}}
\newcommand{\rhob}{\widetilde{\rho}}

\newcommand{\dnu}{\delta \nu}

\newcommand{\nc}{{\cal N}}
\newcommand{\ncb}{\widetilde{\cal N}} 
\newcommand{\nub}{\widetilde{\nu}} 
\newcommand{\nct}{\delta {\cal N}}
\newcommand{\eb}{\widetilde{E}}
\newcommand{\et}{\delta {E}} 
\newcommand{\gu}{g_{\rm{u}}}
\newcommand{\Ai}{{\rm{Ai}}}

\begin{document}

\title [Cold atoms at unitarity and semiclassical ground state ...]
{Cold atoms at unitarity and semiclassical ground state of a Fermi gas for Haldane-Wu exclusion statistics}

\author{J\'er\^ome Roccia$^{1,2}$}

\address{
$^1$Institut de Physique et Chimie des Mat{\'e}riaux de Strasbourg,
CNRS-UdS, UMR 7504, 
23 rue du Loess, BP 43, 67034 Strasbourg Cedex 2, France\\
$^2$Institute for Theoretical Physics, University of
Regensburg, D-93040 Regensburg, Germany}

\date{}

\begin{abstract} 
We investigate finite particle systems of cold atoms bound in a local potential $V(\bfr)$. We derive the ground state energy and the particle density using a recently developed semiclassical theory (2008 {\it Phys. Rev. Lett.} {\bf 100} 200408), and assuming the particles are described by the Haldane-Wu fractional exclusion statistics (FES) at unitarity. This approach is applied to atoms trapped into a three dimensional harmonic oscillator. We show that the parameter-free FES semiclassical theory yields results that are consistent with numerical simulations by Chang and Bertsch [2007 {\it Phys. Rev} A {\bf 76} 021603(R)] and  Bulgac (2007 {\it Phys. Rev.} A {\bf 76} 040502).
\end{abstract}










\noindent

Recently, the crossover between the BEC (Bose-Einstein condensate) and BCS (Bardeen-Cooper-Schrieffer superconductivity) regimes in a cold atom gas has attracted a great deal of attention \cite{regal}. Using the effect of the Feshbach resonances, the interactions or the scattering length $a$ can be tuned by an external magnetic field. When increasing the magnetic field, $a$ grows from a small negative value in the BCS regime to a small positive value in the condensate state. In between these two regimes, the scattering length diverges and the gas is said to be at unitarity. Since no relevant energy scale exists but the Fermi wave length, one expects a universal behavior of the gas \cite{gps}. In that picture Papenbrock \cite{pap} noticed that the ground state energy of this system grows like the Thomas-Fermi (TF) approximation. The proportionality constant, expected to be universal, has been recently computed analytically by Bhaduri {\it et al.} \cite{bms1,bmb,bms2}, assuming interactions at unitarity are simulated by non-interacting particles which obey the fractional Haldane-Wu's statistics (FES) \cite{haldane}.
This extension of standard Bose-Einstein and Fermi-Dirac statistics was computed in connection with the fractional quantum Hall effect, and later extended by Wu \cite{wu}. In FES, the Pauli exclusion principle is generalized such that the occupation number follows $n_g(\epsilon,\lambda)=1/(w[e^{(\epsilon -\lambda)/k_{{\rm B}}T}]+g)$ where $T$ is the temperature, $g$ is the occupancy factor, $\lambda$ is the chemical potential and $w^g(x)[1+w(x)]^{1-g}=x$. We recover the Bose and Fermi distributions, respectively, with $g=0$ and $g=1$. At zero temperature the FES occupation number obeys the step distribution 
\begin{equation}\label{nboc}
n_g(\epsilon,\lambda)= \frac{1}{g} ~\Theta(\epsilon-\lambda) ~,
\end{equation}
where $\Theta(x)$ is the Heaviside function.
For such statistics the semiclassical theory, initiated by the {\it trace formula} of Gutzwiller \cite{gutz} and recently extended in \cite{rb} for spatially-varying densities, provides a convenient formalism to compute the relevant thermodynamic quantities for finite particle systems \cite{mako,olof,rbkm,bjp}.

In this communication, we establish the FES semiclassical theory for the ground state energy and for the particle density. We discuss their expectation values for amplitudes and shell effects. We check to what extent the FES hypothesis of unitary atoms trapped into a three dimensional isotropic harmonic potential (IHO) agrees with benchmark numerical simulations performed by Chang and Bertsch \cite{chang} and Bulgac \cite{bulgac}.




We first derive the semiclassical theory for $N$ non-interacting particles 
with mass $m$, obeying FES statistics and bound by a local smooth
potential $V({\bf r})$ at zero temperature.
Here we restrict ourselves to three-dimensional systems, but the generalization to other dimensions is straightforward.
The discrete energy eigenvalues $\epsilon_j$ and eigenfunctions $\psi_j(\bfr)$ 
are given by the stationary Schr\"odinger equation. Using the FES occupation number (\ref{nboc}), the
quantum-mechanical single-particle (SP) level density and particle density of the system at zero temperature, 
 are given respectively by $\nu(\epsilon)=\sum_{\epsilon_j}\delta (\epsilon-\epsilon_j)$ and
\begin{equation} 
\rho(\bfr)= 2 \sum_{\epsilon_j} n_g(\epsilon_j,\lambda) \psi_j^{\star}({\bf r}) \psi_j(\bfr)	
=\frac{2}{g}\!\sum_{\epsilon_j\leq \lambda } \psi_j^{\star}({\bf r}) \psi_j(\bfr) \, . \label {rho}
\end{equation}
The chemical potential is determined by the integrated level density which counts the number of SP state up to $\lambda$:
\begin{equation}\label{stair}
\nc(\lambda)=2\int_0^\infty n_g(\epsilon,\lambda) \nu(\epsilon) d\epsilon=\frac{2}{g}\int_0^\lambda \nu(\epsilon) d\epsilon .
\end{equation}
The factor 2 in (\ref{rho}) and (\ref{stair}) takes into account the spin degeneracy.
For a given potential the SP level density and the particle density are decomposed into a smooth 
part $\nub$ (resp. $\rhob$) coming from the TF theory plus an oscillatory contribution
$\delta \nu$ (resp. $\delta \rho$):
\begin{eqnarray} 
\nu (\epsilon) &=& {\widetilde{\nu}} (\epsilon) + \dnu (\epsilon) \label{eq101}  ,\\
\rho(\bfr)&=&{\widetilde\rho}(\bfr)+\delta\rho(\bfr)\label{densep}.
\end{eqnarray}
Below, we will show in more 
detail how to describe these two components. (\ref{eq101}) induces a 
similar decomposition for the integrated level density:
\begin{eqnarray}
\nc (\lambda) &=& \ncb (\lambda) + \nct (\lambda) \nonumber\\
&=& \frac{2}{g} \int^{\lambda}_0 \nub(\epsilon) \, d\epsilon \ 
+ \frac{2}{g}\int^{\lambda}_0 \dnu(\epsilon) \, d\epsilon \ .\label{eq102}
\end{eqnarray}
In order to study a system with a finite number of
particles, we use the canonical expressions for
thermodynamic quantities. For a system of $N$ non--interacting fermions, 
we define its ground state energy $E(N)$, the shell correction energy $\et (N)$
and the smooth TF component $\eb (N)$ as \cite{book,rs}:
\begin{eqnarray}
\et(N) &=& E(N)-\eb(N) \nonumber \\
&=& 2 \int_0^{\infty} n_g(\epsilon,\lambda) \epsilon \nu(\epsilon) d\epsilon
- 2 \int_0^{\infty} n_g(\epsilon,\lambdab)\epsilon \nub(\epsilon) d\epsilon \nonumber \\
&=& \frac{2}{g} \int_0^{\lambda} \epsilon \nu(\epsilon) d\epsilon
- \frac{2}{g} \int_0^{\lambdab} \epsilon \nub(\epsilon) d\epsilon .\label{eq103}
\end{eqnarray}
The chemical potential $\lambda$ and its smooth part
$\lambdab$ fix the number of particles.
They are defined by inversion of the exact and the average integrated level densities, 
$\nc (\lambda) = N$ and 
\begin{equation} \label{ncb} 
\ncb (\lambdab) = N \ ,
\end{equation}
respectively. Note that, from (\ref{eq102}) and (\ref{ncb}), both $\lambda$ and $\lambdab$ depend on $g$. Unfortunately, (\ref{eq103}) is difficult to exploit 
analytically because the discretization of
$\lambda$ is difficult to impose. From (\ref{eq103}) it can
be shown that, neglecting terms of second order in the parameter $\lambda - \lambdab$, $\et$ 
may be approximated by \cite{mbc,lm}:
\begin{equation} \label{eq104}
\et(N) \approx  - \int_0^{\lambdab} \nct(\epsilon) d\epsilon \ .
\end{equation}
This, together with the definitions of $\eb (N)$, $\rhob(\bfr)$ and $\delta \rho(\bfr)$, are the basic equations upon which 
we will base our analysis of the ground state of FES Fermi gases.

In the limit $N\rightarrow \infty$, $\rho$ and $E$ are expected to carry over into the approximation obtained in the TF theory \cite{matf}.
${\widetilde\nu}$ is given for any local potential $V(\bfr)$, by
\begin{equation}\label{nub}
{\widetilde\nu} = \nu_{_{\rm{TF}}} = \frac{m^{3/2}}{\pi^2 \hbar^3 \sqrt{2}}
\int\sqrt{\epsilon-V(\bfq)}\Theta[\epsilon-V(\bfq)]d \bfq .
\end{equation}
The TF expression for the particle density has to be weighted by the FES occupation number (\ref{nboc}) following (\ref{rho}).
It yields:
\begin{equation}\label{dnu}
{\widetilde\rho} = \rho_{_{\rm{TF}}} =  \frac{8}{3\sqrt{\pi}g}\left(\frac{m}{2\pi\hbar^2}\right)^{3/2}
[\lambdab-V(\bfr)]^{3/2} ,
\end{equation}
where $\lambdab$ is given by inverting (\ref{ncb}) with the l.h.s. of (\ref{eq102}) and (\ref{nub}). $\eb$ is computed using the r.h.s. of (\ref{eq103}) and (\ref{nub}) :
\begin{equation}\label{eqeb}
{\eb} (N)=\frac{2 m^{3/2}}{\pi^2 \hbar^3 \sqrt{2} g}
\int_0^{\lambdab} \int \epsilon \sqrt{\epsilon-V(\bfq)}\Theta[\epsilon-V(\bfq)]d \bfq d\epsilon.
\end{equation}


For the oscillating parts $\delta \rho$ and $\delta E$, we now use the main 
formulas from \cite{gutz,rb,mbc} for the special case at $D=3$ space dimensions.
Including the FES occupancy factor, the semiclassical 
expression for $\delta \rho$, to leading order in $\hbar$, yields
\begin{equation}\label{drhosc}
\delta \rho(\bfr)   \simeq  \frac{2}{g}\,\sum_\gamma  \frac{m \sqrt{|{\cal D}_{\bot\gamma}|}}{ \pi^2 p T_\gamma}
\,\cos\left(\frac{1}{\hbar}S_\gamma-\mu_\gamma\frac{\pi}{2}-\pi\right).
\end{equation}
The sum is, in general, over all non-periodic orbits $\gamma$ starting and ending in $\bfr$ taken at energy $\lambdab$. 
$S_\gamma$ is the action function $
S_\gamma = \int_{\bfr}^{\bfr'=\bfr} {\bf p}(\lambdab,{\bf q})\cdot d\,{\bf q}\,,
$
where ${\bf p}(\lambdab,\bfq)$ is the classical momentum in the point $\bfq$.
$\mu_\gamma$ is the Morse index 
that counts the number of conjugate points along the orbit \cite{gutz}.  
${\cal D}_{\bot\gamma} =(\partial{p}_{\bot}/ \partial{r'_{\!\bot}})|_{\bfr'=\bfr}$ is the 
stability determinant
calculated from the components $p_{\bot}$ and $r'_{\!\bot}$ {\it transverse} 
to the orbit $\gamma$ of the initial momentum and final coordinate, respectively.
Here $p$ is the modulus of the momentum in $\bfr$ and $T_\gamma={\rm d}S_\gamma(\epsilon,\bfr)/{\rm d}\epsilon|_{\epsilon=\lambdab}$ is the 
running time of the orbit $\gamma$.

To leading order in $\hbar$, the oscillating part $\delta \nu(\epsilon)$ is given by the {\it semiclassical trace formula}
\begin{equation}
\delta \nu(\epsilon) \simeq \sum_{_{\rm{PO}}} {\cal B}_{_{\rm{PO}}}(\epsilon) \cos\left[
                   \frac{1}{\hbar}\,S_{_{\rm{PO}}}(\epsilon)-\frac{\pi}{2}\,\sigma_{_{\rm{PO}}}\right], 
\label{trf}
\end{equation}
where the sum runs over all {\it periodic orbits} (POs). For systems in
which all orbits are isolated in phase space, Gutzwiller \cite{gutz} 
derived explicit expressions for the smooth amplitudes ${\cal B}_{_{\rm{PO}}}(\epsilon)$,
which depend on the stability of the orbits, and for the Maslov
indices $\sigma_{_{\rm{PO}}}$. Performing the trace integral in the semiclassical Green function \cite{gutz}
along all directions transverse to each orbit $\gamma$, the stationary 
phase approximation (SPA) leads immediately to the periodicity of the 
contributing orbits. The Maslov index $\sigma_{_{\rm{PO}}}$ collects all 
phases occurring in the semiclassical Green function
and in the SPA for the trace integral 
(see \cite{masl} for detailed computations of $\sigma_{_{\rm{PO}}}$). 
$S_{_{\rm{PO}}}(\epsilon)$ is the closed action integral
$S_{_{\rm{PO}}}(\epsilon) = \oint_{_{\rm{PO}}} \bfp(\epsilon,{\bf q})\cdot d {\bf q}$.
We compute analytically $\delta E$ using the r.h.s. of (\ref{eq102}) and (\ref{eq104}). To leading order in $\hbar$ we get
\begin{equation}\label{desc}
\delta E(N)=\frac{2\hbar^2}{g}\sum_{_{\rm{PO}}} \frac{{\cal B}_{_{\rm{PO}}}(\lambdab)}{T^2_{_{\rm{PO}}}(\lambdab)} \cos\left[
         \frac{1}{\hbar}\,S_{_{\rm{PO}}}(\lambdab)-\frac{\pi}{2}\,\sigma_{_{\rm{PO}}}\right],
\end{equation}
where $T_{_{\rm{PO}}}$ is the running time of the PO.\\
We emphasize that the semiclassical approximations
are not valid in regions close to the classical turning points
$\bfr_\lambda$ defined by $V(\bfr_\lambda)=\lambdab$. Since the
classical momentum $p$
becomes zero, the spatial density (\ref{drhosc}) always
diverges at the turning points. Furthermore the running time which appears in the denominator of (\ref{drhosc}), may vanish at the
turning point for certain orbits. These divergences
can be overcome by linearizing the smooth potential
$V(\bfr)$ around the classical turning points \cite{wkb,circ}.
Note that the phases and amplitudes in (\ref{drhosc}) and (\ref{desc}) strongly depend on $g$ through $\lambdab$, but no simple behavior clearly emerges.
In practice, one has to compute them explicitly for a given potential. In the following, we address this issue in the special case of a 3D IHO applied to a unitary Fermi gas. 


Bhaduri {\it et al} \cite{bms1,bmb,bms2} assume that FES simulates interactions between particles at unitarity and give the analytical expression of the unitary occupancy factor $\gu=1-\sqrt{2}/2\approx0.29$ for a 3D IHO confinement $V(\bfr)=m \omega^2 r^2/2$ with $|\bfr|=r$. They compute
observables in the extended TF limit leading to smooth variations only. Generalizing their method, we introduce oscillating corrections that depend on $\hbar$ for $g=\gu$.
For a 3D IHO, the smooth functions $\rhob$ and $\eb$ are easily computed and give
\begin{eqnarray}
{\widetilde\rho} (r)&=& \frac{8}{3\sqrt{\pi}\gu}\left(\frac{m}{2\pi\hbar^2}\right)^{3/2}
(\lambdab-m\omega^2 r^2/2)^{3/2} \label{dnuho} , \\
 \eb(N)&=&\frac{\hbar \omega}{\gu}\bigg[\frac{1}{4}\bigg (\frac{\lambdab}{\hbar \omega}\bigg)^{4}+\frac{1}{8}\bigg (\frac{\lambdab}{\hbar \omega}\bigg)^{2}+o(1)\bigg], \label{etho}
\end{eqnarray}
with $\lambdab(N)=\hbar \omega (3 \gu N)^{1/3}$.
These two results are consistent with those in \cite{bmb,zyl}.
Using the formulas in \cite{rbkm,jl} for $\delta \rho$ and $\delta E$ with the FES modifications (\ref{drhosc}) and (\ref{desc}), we get:
\begin{equation}\label{drho}
\delta\rho(r) =  
\frac{-m^2\omega(3N)^{1/3}}{\pi^2  r p^2 \gu^{2/3}}\sum_{k=0,\pm}^\infty\!\! \frac{\cos\!\left(S_{\pm}^{(k)}
-\mu_{\pm}^{(k)}\frac{\pi}{2}\right)}{T_{\pm}^{(k)}}~,
\end{equation}

\begin{equation}\label{deho}
\delta E(N)=\frac{\hbar \omega}{\gu^{1/3}}\frac{(3N)^{2/3}}{2\pi^2}\sum_{k=1}^{\infty} \frac{(-1)^k}{k^2}\cos[2\pi k(3\gu N)^{1/3}]~.
\end{equation}

Here we have used the analytical form of the actions and periods
$S_{\pm}^{(k)}  =  (2k+1)\pi\lambdab/\omega \mp rp \mp 2\lambdab/\omega\arctan(m\omega r /p)$,
$T_{\pm}^{(k)}  =  (2k+1) \pi/\omega \mp 2/\omega \arctan(m\omega r/p)$,
$\mu_{+}^{(k)}=6 k +1$ and $\mu_{-}^{(k)}=6 k  +3$.
From (\ref{deho}) we note that the amplitude of the shell correction energy for a unitary Fermi gas is larger ($\delta E/E\propto \gu^{-2/3}$) compare to the standard Fermi gas ($g=1$). The approximate frequency of the shell fluctuations is given by the phase of the cosine function of the $k=1$ term in the sum (\ref{deho}). Thus, the closed-shell (resp. mid-shell) numbers, given by the values of $N$ that minimize (resp. maximize) $\delta E(N)$, are well approximated by $N_{{\rm cs}}= i^3/(3\gu)$ [resp. $N_{{\rm ms}}= (i+1/2)^3/(3\gu)$] with $i \in  \mathbb{N}$. 
The presence of $p$ in the denominator of (\ref{drho}) gives no simple behavior for $\delta \rho$ in $\gu$. Nevertheless close to $r\approx0$, the ratio $\delta \rho/\rhob$ grows like $\gu^{-1/2}$, so that the oscillations are amplified by a factor $1.9$.
\begin{figure}[th]
\begin{center}
\begin{minipage}{1.\linewidth}
\hspace{2.7cm}
\includegraphics[width=.7\columnwidth,clip=true]{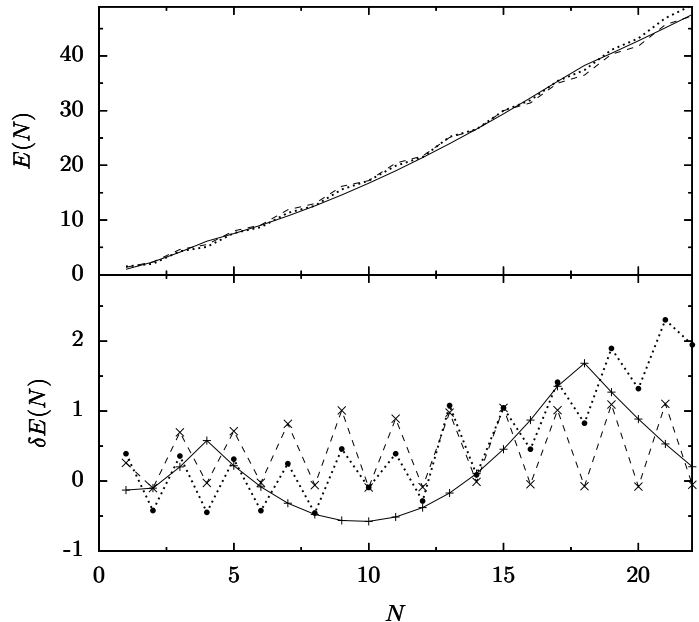} \vspace*{-0.5cm}
\end{minipage}
\end{center}
\caption{\label{energy}
Upper panel: ground state energy of a Fermi gas at unitarity as a function of the particle number $N$ in a 3D IHO. The solid line corresponds to the semiclassical FES ground state energy (\ref{etho})$+$(\ref{deho}), the dotted (resp. dashed) line corresponds to the numerical data \cite{chang} (resp. \cite{bulgac}) (units $\hbar=m=\omega=1$).
Lower panel:  shell correction energy for the same system. The solid line corresponds to the semiclassical FES theory (\ref{deho}). The dotted (resp. dashed) line is the numerical data \cite{chang} (resp. \cite{bulgac}) subtracted by the smooth energy (\ref{etho}).
}
\end{figure}

When $r$ is close to the classical turning point $r_{\lambda}=\omega^{-1}(2 \lambdab/m)^{1/2}$, the semiclassical approximation (\ref{drho}) diverges. Following the regularisation method detailed in \cite{circ}, the density profile becomes
\begin{equation}\label{rhoregul}
\rho(r)\mathop {=}\limits_{r\rightarrow r_{\lambda}}\frac{\rho_0^3}{48\pi \gu} \{\Ai(z)\Ai'(z)+2 z[\Ai'(z)]^2-2 z^2 \Ai^2(z)\}\, ,
\end{equation}
where 
$\rho_0=2 (2 m \omega/\hbar^2)^{1/3} (2 m \lambdab)^{1/6}$, $z=\rho_0(r-r_{\lambda}/2)/2$, $\Ai(x)$ is the Airy function and $\Ai'(x)$ its derivative.

\begin{figure}[ht]
\begin{center}
\begin{minipage}{1.\linewidth}
\hspace{2.7cm}
\includegraphics[width=0.7\columnwidth,clip=true]{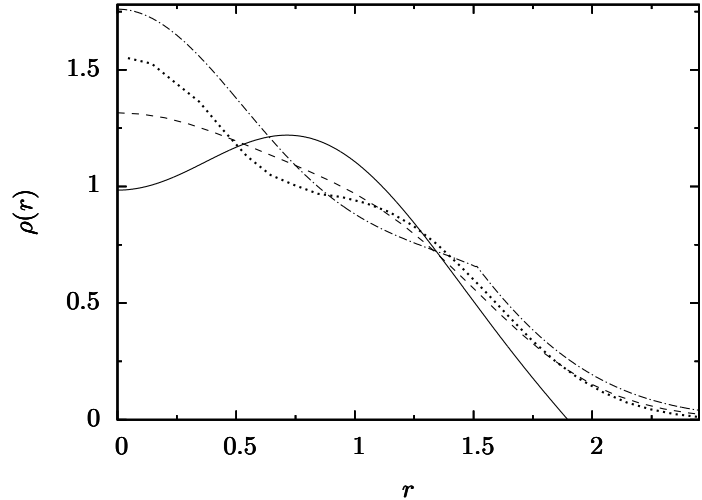} 
\end{minipage}
\end{center}
\caption{\label{densities}
Particle density of a Fermi gas at unitarity as a function of $r$ for $N=20$ particles in a 3D IHO. The solid line corresponds to the semiclassical FES particle density (\ref{dnuho})$+$(\ref{drho}), the dotted (resp. dashed) line corresponds to the numerical data \cite{chang} (resp. \cite{bulgac}) (units $\hbar=m=\omega=1$). The dotted-dashed line is the semiclassical improvement discussed in the text.
}
\end{figure}
We now compare ours results to the {\it ab initio} Green function Monte Carlo method (GFMC) from Chang and Bertsch \cite{chang} and the superfluid local density approximation (SLDA) computed by Bulgac \cite{bulgac}.
Figure \ref{energy} focuses on the ground state energy for $N=2$--$22$ atoms. The upper panel shows a qualitative agreement between the three curves. In the lower panel only the oscillating component is plotted. While the shell effects are observed for the FES model (solid line), they are not present in the SLDA model (dashed line). The reason is LDA theories correspond to (extended) TF approximations \cite{dft}. They only provide the average part of thermodynamic quantities, but cannot take into account shell effects. The GFMC data (dotted line) shows some irregularities but no clear oscillations. This is not surprising since the semiclassical theories are known to be more accurate for large particle numbers. Hence further numerical data are needed in order to check the presence of shell effects at unitarity. Note the odd-even oscillation of the energy in the numerical computations, which is due to the pairing correlations not included in the FES theory.

In figure \ref{densities}, we plot the particle density for $N=20$.
The GFMC (dotted line) and FES (solid line) models show some density oscillations. Although the order of magnitude is correct, the oscillations are not in phase.
This is attributed to the discrepancy of the semiclassical theory for particle numbers far from the shell or mid-shell closure, thus leading to an inaccurate sign of $\delta \rho$ (see \cite{circ} for an exhaustive discussion). The FES model is expected to give best results for $N\approx N_{{\rm cs}}$ and $N\approx N_{{\rm ms}}$. The SLDA model (dashed line) gives an accurate average profile of the particle density but shows no oscillations. We mention that same comparisons have been done with the numerical work of von Stecher {\it et al.} \cite{sgb1,bsg,sgb2,db} leading to similar results with the SLDA model. 
The linear behavior of the FES particle density observed for $r\geq1.5$, is a consequence of the breakdown of the semiclassical approximation near the classical turning point. We recover the usual exponential tail using the regularised formula (\ref{rhoregul}).
The dotted-dashed line illustrates an improved density profile for which we have changed the overall sign into (\ref{drho}) and switched to the regularised expression (\ref{rhoregul}) for $r\geq1.5$: in this case, the agreement with GFMC is much better.

In conclusion, we have investigated finite fermions systems which are described by the Haldane-Wu statistics going beyond the TF approximation.
The ground state energy and the particle density of this system are derived analytically at zero temperature. We used the FES semiclassical theory as a parameter-free model of unitary Fermi gases and we discussed shell effects as a function of the particle number and the position. 
Considering that the semiclassical model is expected to give better results for large $N$, 
we gave reasonably good agreement with two numerical studies. The investigation of the more general occupation number distribution $n_g(\epsilon,\lambda)=1/(w[e^{(\epsilon -\lambda)/k_{\rm{B}}T}]+g)$ in the case of non-zero temperatures is let for future works.


I acknowledge A Bulgac, S Y Chang and D Blume for providing the numerical data and M Brack for fruitful discussions and constant encouragement. This work was funded in part by the French National Research Agency ANR (project ANR-06-BLAN-0059).

\end{document}